\documentclass[twocolumn,showpacs,preprintnumbers,amsmath,amssymb,amsmath,aps,prb,lengthcheck]{revtex4-1}
\usepackage{amssymb}
\usepackage{graphicx}
\usepackage{dcolumn}
\usepackage{bm}
\usepackage{hyperref}
\usepackage{makeidx}
\usepackage{bbm}
\usepackage{wasysym}

\usepackage[english]{babel}
\usepackage[latin1]{inputenc}
\usepackage{color}
\usepackage{graphicx}
\usepackage{amsmath}

\usepackage{latexsym}
\usepackage{amsthm}

\makeindex

\def\>{\right\rangle}
\def\<{\left\langle}
\def\be{\begin{equation}}
\def\ee{\end{equation}}
\def\ba{\begin{array}{l}}
\def\ea{\end{array}}

\def\beq{\begin{eqnarray}}
\def\eeq{\end{eqnarray}}

\begin{document}

\preprint{APS/123-QED}

\title{Current enhancement through a time dependent constriction in fractional topological insulators}

\author{G. Dolcetto,$^{1,2,3}$ L. Vannucci,$^{1}$ A. Braggio,$^{2,3}$ R. Raimondi,$^{4}$ M. Sassetti,$^{1, 2}$}
\affiliation{$^1$ Dipartimento di Fisica, Universit\` a di Genova, Via Dodecaneso 33, 16146, Genova, Italy.\\
$^2$ CNR-SPIN, Via Dodecaneso 33, 16146, Genova, Italy.\\
$^3$ INFN, Via Dodecaneso 33, 16146, Genova, Italy.\\
$^4$ Dipartimento di Matematica e Fisica, Universit\` a Roma Tre,
Via della Vasca Navale 84, Rome, Italy,\\}
\date{\today}

\begin{abstract}
We analyze the backscattering current induced by a time dependent constriction as a tool to probe fractional topological insulators.
We demonstrate an enhancement of the total current for a fractional topological insulator induced by the dominant tunneling excitation, contrary to the decreasing present in the integer case for not too strong interactions.
This feature allows to unambiguously identify fractional quasiparticles.
Furthermore, the dominant tunneling processes, which may involve one or two quasiparticles depending on the interactions, can be clearly determined.
\end{abstract}

\maketitle

\section{Introduction}\label{Intro}
The close interplay between theory and experiments has always led to fundamental discoveries in condensed matter physics.
This is particularly evident in the search for topological states of matter, starting in the 80's with the experimental observation of the integer quantum Hall (QH) effect \cite{vonKlitzing80}, characterized by chiral metallic edge states and insulating bulk states.
Few years later, also fractional plateau of the conductance were discovered \cite{Tsui82}, theoretically explained by Laughlin as a manifestation of the fractional QH effect \cite{Laughlin83}.
After the first theoretical proposal for realizing topological states of matter in the absence of magnetic fields \cite{Haldane88, Kane05}, Bernevig \textit{et al.} suggested that HgTe/CdTe quantum wells behave as a topological insulator (TI) \cite{Hasan10, Qi11}. realizing the quantum spin Hall (QSH) effect in the presence of time reversal (TR) symmetry \cite{BernevigNat06}. Soon after the group of Molenkamp provided the first experimental report \cite{Konig07}, showing that nonlocal transport is due to protected helical edge states \cite{Roth09}, with spin and momentum bound to each other \cite{BernevigPRL06}.
With this picture in mind, it is natural to wonder how long the experimental realization of fractional TIs, recently proposed by Levin and Stern \cite{Levin09, Levin12}, will take.
At the simple intuitive level, fractional TIs can be thought as a superposition of two copies of fractional QH states in opposite magnetic fields, with the emergence of TR protected helical edge states with fractional excitations.
Beyond the emergence of fractional QSH effect in the flat-band lattice models \cite{NeupertPRB11, NeupertPRL12, Li14}, possible experimental realization of such a state have been proposed, ranging from a fractional QH state with a varying $g$-factor in the presence of a thin insulating barrier to electron-hole bilayers realized in electron-hole symmetric systems \cite{Lindner12}.\\
It is thus important to find possible strategies and experimental signatures to identify this novel topological state of matter, which could be exploited also for quantum computing \cite{Cheng12, Burrello13, Nikolic13}.
A breakthrough in this direction has been done recently by Beri and Cooper \cite{Beri12}, who demonstrated a surprising robustness of fractional TIs to magnetic perturbations, which is completely unexpected in integer TIs.
Based on the works carried out in fractional QH systems and integer TIs \cite{dePicciotto97, Saminadayar97, Chung03, Dolev08, Carrega11, Teo09, Hou09, Dolcetto12,Schmidt11, Liu11, Strom09}, they also showed peculiar transport properties when electrons tunnel through a quantum point contact (QPC), which could be used to probe fractional TIs \cite{Beri12, Huang13}.\\
In this context, we propose tunneling through a time dependent QPC, realized, for example, by applying time modulated gate voltages, to probe fractional TIs.
In the case of a static QPC, backscattering at the constriction always decreases the total current, and information about the tunneling processes are encoded in the power-law exponents as a function of bias and temperature.
However, the power laws are often hardly detectable and hidden by several additional mechanisms.
Here, we propose a different and more robust method in order to discriminate between fractional and integer TIs.
We demonstrate that in fractional TIs a time dependent QPC enhances the current, contrary to the decreasing induced in integer TIs for not too strong interactions.
This peculiar effect is strictly associated with quasiparticle tunneling, and provides a simple tool to identify fractional TIs.
This method takes advantage from the resolving power of finite frequency transport properties in the simple framework of dc current measurements.
Different tunneling processes are present, involving one or more excitations depending on the strength of interactions; we show that the proposed setup also provides a simple way to discriminate between these different processes, without requiring the knowledge of the detailed form of the current.\\
The paper is organized as follows. In Sec. \ref{Model} we describe the interacting helical edge states of fractional TIs and define the relevant tunneling processes.
In Sec. \ref{bscurrent} we evaluate the backscattering current induced by the time dependent QPC in the weak backscattering regime, and discuss how it is affected by the different tunneling processes, both in integer and in fractional TIs.
This analysis leads us to draw the main conclusions of our work.
Finally, Sec. \ref{conclusions} is devoted to the conclusions.

\section{Model}\label{Model}
\begin{figure}[!ht]
\centering
\includegraphics[width=8cm,keepaspectratio]{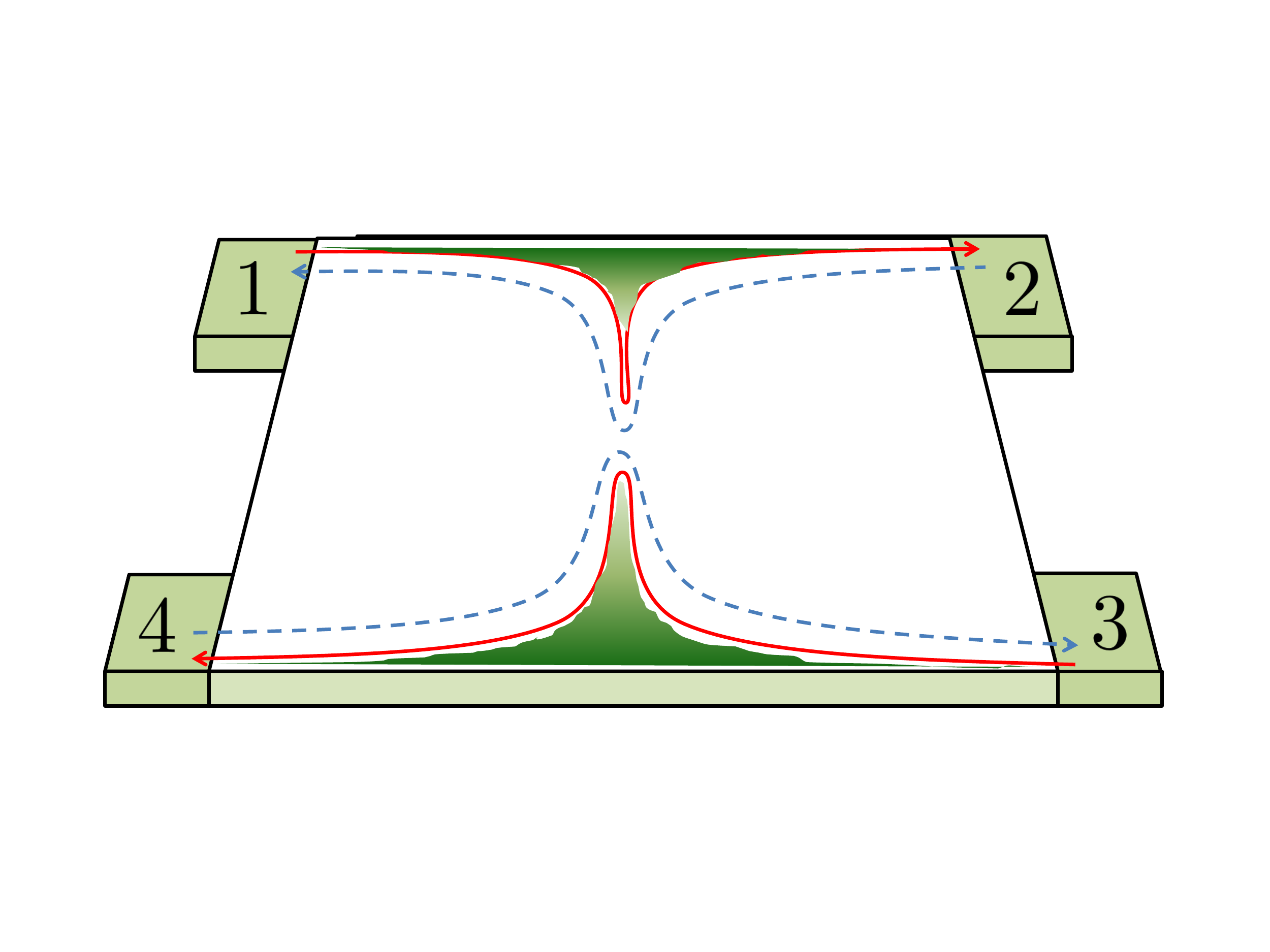}
\caption{(Color online) Four-terminal setup for the fractional TI with a time dependent QPC. Solid red (dashed blue) lines represent spin up (down) excitations.}\label{Fig:4term}
\end{figure}

\begin{figure}[!ht]
\centering
\includegraphics[width=7cm,keepaspectratio]{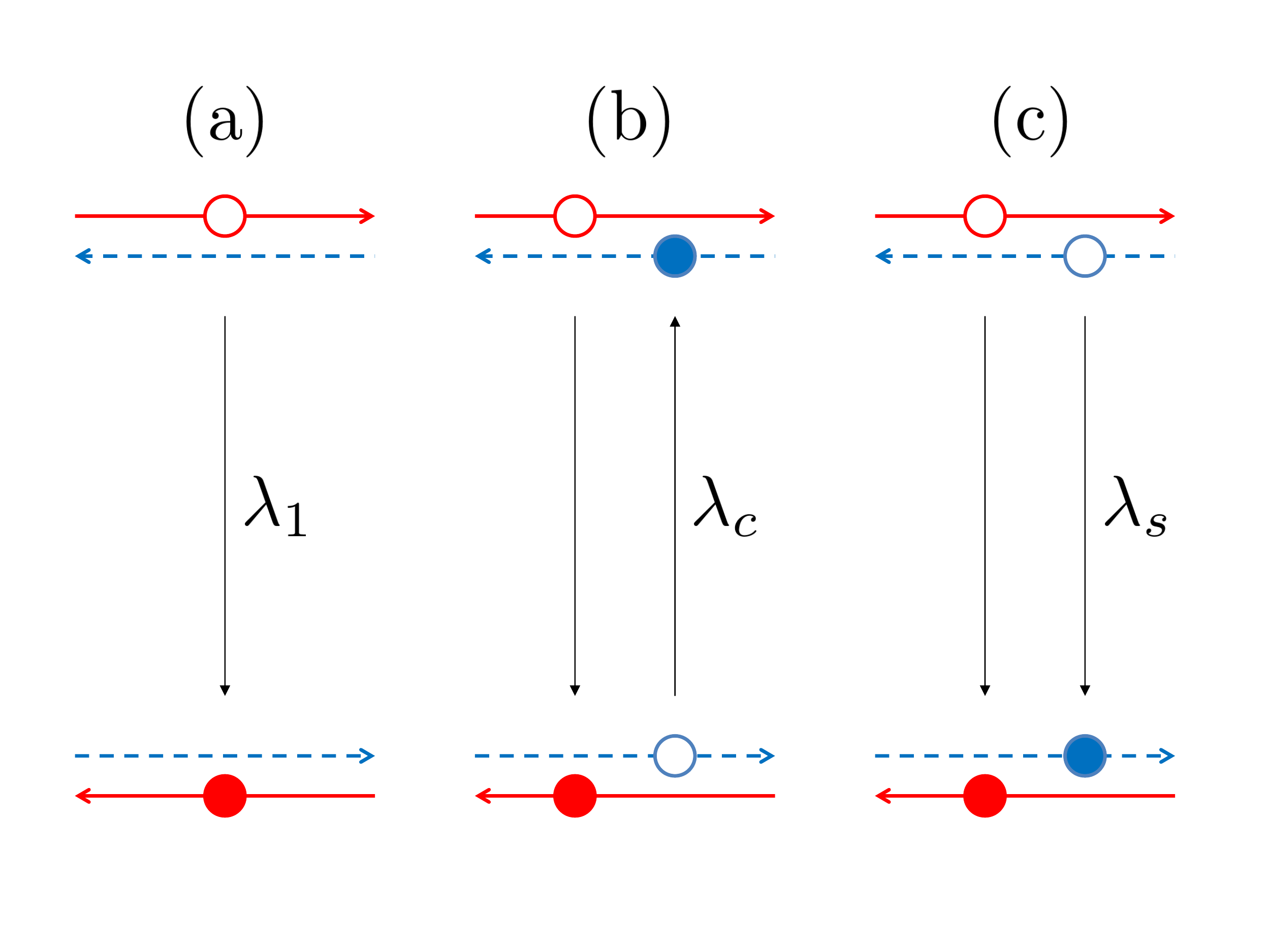}
\caption{Some examples of the dominant tunneling processes in the weak backscattering regime: (a) single quasiparticle, (b) charge and (c) spin tunneling, with amplitudes $\lambda_1$, $\lambda_c$ and $\lambda_s$ respectively. Solid red (dashed blue) lines represent spin up (down) excitations.}\label{Fig:processes}
\end{figure}

We consider the upper edge of the fractional QSH bar with right-moving spin up ($R\uparrow$) and left-moving spin down ($L\downarrow$) excitations, and the opposite on the lower edge.
The effective Lagrangian density is \cite{Levin09, Levin12} ($\hbar=1$)
\begin{equation}\label{eq:Lagrangian}
\mathcal{L}_0=-\frac{1}{4\pi\nu}\sum_{\alpha=R,L}\sum_{\sigma=\uparrow,\downarrow} \partial_x\phi_{\alpha\sigma}\left (\alpha\partial_t+v\partial_x\right )\phi_{\alpha\sigma},
\end{equation}
with $v$ the propagation velocity, $\alpha=R/L\equiv+/-$ and $\sigma=\uparrow/\downarrow$.
The filling factor is $\nu=1/(2n+1)$, with $n=0$ for integer TIs and $n\in \mathbb{N}_0$ for fractional TIs in the Laughlin sequence.
The bosonic fields $\phi_{\alpha\sigma}$ obey the commutation relations $[\phi_{\alpha\sigma}(x),\phi_{\alpha '\sigma '}(x')]=i\alpha\nu\pi\delta_{\alpha,\alpha '}\delta_{\sigma,\sigma '}\mathrm{sign}(x-x')$.
They are related to electron $\Psi_{\alpha\sigma}$ and quasiparticle $\psi_{\alpha\sigma}$ operators by
\begin{equation}\label{eq:operators}
\Psi_{\alpha\sigma}=\frac{e^{\frac{i}{\nu}\alpha\phi_{\alpha\sigma}}}{\sqrt{2\pi a}}, \ \ \ \ \ \
\psi_{\alpha\sigma}=\frac{e^{i\alpha\phi_{\alpha\sigma}}}{\sqrt{2\pi a}},
\end{equation}
with $a$ the microscopic cutoff length, and respectively correspond to excitations with charge $-e$ and $-e^*=-\nu e$. We have omitted the Klein factors and the phase factors $e^{i\alpha k_F x}$, which are not relevant for our discussion.
Note that the two operators in Eq. \eqref{eq:operators} coincide in the case of integer TIs, since the fundamental excitation is the electron itself.
The Hamiltonian is easily obtained from Eq. \eqref{eq:Lagrangian}
\begin{equation}\nonumber
H_0=\frac{v}{4\pi\nu}\sum_{\alpha=R,L}\sum_{\sigma=\uparrow,\downarrow}\int dx\left (\partial_x\phi_{\alpha\sigma}\right )^2.
\end{equation}
The presence of electron-electron interaction on the edges is due to intra-mode interaction
\begin{equation}\label{eq:Hee_f}
H_4=\sum_{\alpha=R,L}\sum_{\sigma=\uparrow,\downarrow}\int dx~dx'~\rho_{\alpha\sigma}(x)V_4(x-x')\rho_{\alpha\sigma}(x')
\end{equation}
and inter-modes interaction
\begin{equation}\label{eq:Hee_b}
H_2=\sum_{\alpha=R,L}\sum_{\sigma=\uparrow,\downarrow}\int dx~dx'~\rho_{\alpha\sigma}(x)V_2(x-x')\rho_{\bar{\alpha}\bar{\sigma}}(x')
,\end{equation}
with $\rho_{\alpha\sigma}(x)=:\Psi_{\alpha\sigma}^{\dagger}(x)\Psi_{\alpha\sigma}(x):=\frac{1}{2\pi}\partial_x\phi_{\alpha\sigma}(x)$ the electron density associated to the mode with chirality $\alpha$ and spin $\sigma$.
Note that not all the terms appear in Eq. \eqref{eq:Hee_b}, since modes with same chirality and different spin are spatially separated, so that we can neglect their mutual interaction.
In the following we consider local interactions $V_{2(4)}(x-x')\approx g_{2(4)}\delta(x-x')$.
The Hamiltonian $H=H_0+H_4+H_2$ can be diagonalized by introducing charge and spin fields \cite{Giamarchi}
\begin{equation}\nonumber
\phi_{\alpha\sigma}=\frac{1}{\sqrt{2}}\left (\nu\varphi_c+\alpha\theta_c+\sigma\varphi_s+\alpha\sigma\nu\theta_s\right ),
\end{equation}
which satisfy $[\varphi_{\lambda}(x),\theta_{\lambda '}(x')]=i\frac{\pi}{2}\delta_{\lambda,\lambda '}\mathrm{sign}(x-x')$,
so that it takes the form of a spinful Luttinger liquid \cite{Giamarchi}
\begin{equation}\nonumber
H=\frac{u}{2\pi}\sum_{\lambda=c,s}\int dx\left [\tilde{K}_{\lambda}(\partial_x\theta_{\lambda})^2+\frac{1}{\tilde{K}_{\lambda}}(\partial_x\varphi_{\lambda})^2\right ].
\end{equation}
The charge and spin parameters are $\tilde{K}_c=K/\nu$ and $\tilde{K}_s=1/\tilde{K}_c$, with
\begin{equation}\nonumber
K=\sqrt{\frac{1+\frac{\nu g_4}{\pi v}-\frac{\nu g_2}{\pi v}}{1+\frac{\nu g_4}{\pi v}+\frac{\nu g_2}{\pi v}}}
\end{equation}
and renormalized velocity
\begin{equation}\nonumber
u=v\sqrt{\left (1+\frac{\nu g_4}{\pi v}\right )^2-\left (\frac{\nu g_2}{\pi v}\right )^2}.
\end{equation}
Here we focus on standard repulsive interactions, which imply $K<1$ ($K=1$ in the absence of interactions) \cite{ref2}.\\
In the presence of a constriction, tunneling from one edge to the other is allowed.
In this work we focus on the weak backscattering regime, where tunneling is a small perturbation to the decoupled edges configuration, focusing on the four terminal setup depicted in Fig. \ref{Fig:4term}.
In this regime tunneling of quasiparticles dominate over electron tunneling for any interaction \cite{Huang13}.
The most relevant processes are sketched in Fig. \ref{Fig:processes}. They correspond to tunneling of single quasiparticles ($H_1$, Fig. \ref{Fig:processes}(a)), backscattering of two quasiparticles ($H_c$, Fig. \ref{Fig:processes}(b)) and tunneling of two quasiparticles from one edge to the other ($H_s$, Fig. \ref{Fig:processes}(c)). Note that $H_{c(s)}$ preserves the charge (spin) on each edge, thus one refers to it as the charge (spin) process.
The corresponding Hamiltonians, assuming without loss of generality the QPC located at $x=0$, are
\begin{eqnarray}
H_1&=&\lambda_1\left [\psi^{\dagger}_{L\uparrow}\psi_{R\uparrow}+\psi^{\dagger}_{L\downarrow}\psi_{R\downarrow}\right ]+\mathrm{h.c.}\nonumber,\\
H_c&=&2\pi a\lambda_c\left [\psi^{\dagger}_{L\uparrow}\psi^{\dagger}_{L\downarrow}\psi_{R\uparrow}\psi_{R\downarrow}\right ]+\mathrm{h.c.}\nonumber,\\
H_s&=&2\pi a \lambda_s\left [\psi^{\dagger}_{L\uparrow}\psi^{\dagger}_{R\downarrow}\psi_{R\uparrow}\psi_{L\downarrow}\right ]+\mathrm{h.c.}\nonumber,
\end{eqnarray}
where $\lambda_p$, with $p=1,c,s$, are the tunneling amplitudes.
In the following we consider a time dependent constriction, with $\lambda_p\to \lambda_p\cos(\omega t)$. This can be achieved through a periodic modulation of the gate voltages that define the QPC.\\
An additional time dependence is introduced by gauging away \cite{Feldman} the bias voltages $V_i$, with $i$ labeling the four terminals (see Fig. \ref{Fig:4term}).
Finally the bosonized tunneling Hamiltonians read
\begin{eqnarray}
H_1&=&\frac{2\lambda_1}{\pi a}\cos(\omega t)\cos\left (\sqrt{2}\nu\varphi_c-e^*V_ct\right )\nonumber\\
&\times&\cos\left (\sqrt{2}\varphi_s-e^*V_st\right ),\label{eq:boso1}\\
H_c&=&\frac{\lambda_c}{\pi a}\cos(\omega t)\cos\left (2\sqrt{2}\nu\varphi_c-2e^*V_ct\right ),\label{eq:bosoc}\\
H_s&=&\frac{\lambda_s}{\pi a}\cos(\omega t)\cos\left (2\sqrt{2}\varphi_s-2e^*V_st\right ).\label{eq:bosos}
\end{eqnarray}
From Eqs. \eqref{eq:boso1}, \eqref{eq:bosoc} and \eqref{eq:bosos} we note that only two combinations of the bias voltages are relevant: the charge
\begin{equation}\nonumber
V_c=\frac{1}{2}\left (V_1-V_2-V_3+V_4\right )
\end{equation}
and spin bias
\begin{equation}\nonumber
V_s=\frac{1}{2}\left (V_1+V_2-V_3-V_4\right ),
\end{equation}
which couple to the charge and spin sectors respectively.
In Eq. \eqref{eq:boso1} the single quasiparticle charge couples to both the charge and spin bias voltages, while in Eq. \eqref{eq:bosoc} (Eq. \eqref{eq:bosos}) a double charge couples to the charge (spin) bias only.\\
The charge and spin bias voltages are also related to the charge ($I_c^{(0)}$) and spin ($I_s^{(0)}$) currents flowing through the topological bar in the absence of the QPC, $I^{(0)}_{c(s)}=G_{c(s)}V_{c(s)}$, with $G_c=2eG_s=\nu e^2/\pi$ the charge and spin conductances.
\section{Backscattering current}\label{bscurrent}
The total charge current in the presence of the constriction is $I_c^{(tot)}=I_c^{(0)}+I_c^{(bs)}$, with $I_c^{(bs)}$ the backscattering current induced by tunneling at the constriction
\begin{equation}\nonumber
I_c^{(bs)}=e^*\frac{dN_R}{dt}=-ie^*\left [N_R,H_1+H_c+H_s\right ].
\end{equation}
Here, $N_R=\sum_{\sigma=\uparrow,\downarrow}\int dx~ \psi^{\dagger}_{R\sigma}\psi_{R\sigma}$ is the total number of right-moving quasiparticles.
The backscattering current can be written as $I_c^{(bs)}=\sum_{p=1,c}I_{p}$, where different contributions are due to the different tunneling processes $I_{p}=-ie^*\left [N_R,H_p\right ]$, with
\begin{eqnarray}
I_{1}(t)&=&-e^*\frac{2\lambda_1}{\pi a}\cos(\omega t)\cos\left (\sqrt{2}\varphi_s-e^*V_st\right )\nonumber\\
&\times&\sin\left (\sqrt{2}\nu\varphi_c-e^*V_ct\right )\nonumber,\\
I_{c}(t)&=&-e^*\frac{2\lambda_c}{\pi a}\cos(\omega t)\sin\left (2\sqrt{2}\nu\varphi_c-2e^*V_ct\right ).\nonumber
\end{eqnarray}
Note that the spin process does not contribute to the backscattering, since $H_s$ preserves the total number of right-moving excitations \cite{Souquet12, Lee12}.\\
We begin focusing on the charge configuration with $V_c\neq 0$ and $V_s=0$, corresponding to a standard two terminal setup with $V_1=V_4=V/2$ and $V_2=V_3=-V/2$. We will comment on alternative configurations in the last part of the work.
We employ the Keldysh formalism \cite{Martin} in order to evaluate the expectation value of the backscattering currents at lowest order in the tunneling amplitudes
\begin{eqnarray}
&&\langle I_{p}(t)\rangle=\frac{1}{2}\sum_{\eta=\pm}\left\langle T_K I_{p}(t^{\eta}) e^{-i\int_{c_K}dt_1H_p(t_1)}\right\rangle\nonumber\\
&=&-\frac{i}{2}\sum_{\eta,\eta_1=\pm}\eta_1\int_{-\infty}^{\infty}dt_1\left \langle T_KI_p(t^{\eta})H_p(t_1^{\eta_1})\right \rangle+O(\lambda_p^4)\nonumber
,\end{eqnarray}
with $T_K$ the Keldysh time-ordering and $\eta=\pm$ labeling the two branches of the Keldysh contour $c_K$.
The time dependent QPC generates dc and ac contributions at frequency $2\omega$. In the following we will focus on the dc components only, which read
\begin{eqnarray}
\langle I_1^{(\omega)}(V_c)\rangle&=&-i2e^*\left |\tilde{\lambda_1}\right |^2\sum_{\eta,\eta_1}\eta_1\int_{-\infty}^{\infty} d\tau \cos\left (\omega\tau\right )\nonumber\\
&\times&\sin\left (e^*V_c\tau\right )e^{2\nu\left [\nu\mathcal{G}_c^{\eta,\eta_1}(\tau)+\frac{1}{\nu}\mathcal{G}_s^{\eta,\eta_1}(\tau)\right ]},\label{eq:I1K}\\
\langle I_c^{(\omega)}(V_c)\rangle&=&-ie^*\left |\tilde{\lambda_c}\right |^2\sum_{\eta,\eta_1}\eta_1\int_{-\infty}^{\infty} d\tau \cos\left (\omega\tau\right )\nonumber\\
&\times&\sin\left (2e^*V_c\tau\right )e^{8\nu^2\mathcal{G}_c^{\eta,\eta_1}(\tau)}\label{eq:IcK}
,\end{eqnarray}
with $2\pi a\tilde{\lambda}_p=\lambda_p$. For clarity, we explicitly write the dependence on $\omega$ and $V_c$ in the dc component.
Since the contributions in Eqs. \eqref{eq:I1K}, \eqref{eq:IcK} are even (odd) functions of $\omega$ ($V_c$), in the following we restrict the discussion to the case $\omega,V_c>0$.
Because of the parity properties of the Keldysh Green's functions $\mathcal{G}_{\lambda}^{\eta,\eta_1}(t-t_1)=\langle \varphi_{\lambda}(t^{\eta})\varphi_{\lambda}(t_1^{\eta_1})\rangle -\frac{1}{2}\langle\varphi_{\lambda}(t^{\eta})\varphi_{\lambda}(t^{\eta})\rangle -\frac{1}{2}\langle\varphi_{\lambda}(t_1^{\eta_1})\varphi_{\lambda}(t_1^{\eta_1})\rangle$, Eqs. \eqref{eq:I1K} and \eqref{eq:IcK} can be written in terms of $\mathcal{G}_{\lambda}^{-,+}(\tau)=\frac{\tilde{K}_{\lambda}}{2}\ln f(\tau)$ only, with
\begin{eqnarray}\nonumber
f(\tau)=\frac{\left |\Gamma\left (1+\frac{1}{\beta\omega_c}-i\frac{\tau}{\beta}\right )\right |^2}{\Gamma^2\left (1+\frac{1}{\beta\omega_c}\right )(1+i\omega_c\tau)}
,\end{eqnarray}
$\Gamma(x)$ being the Gamma function, $\beta$ the inverse temperature and $\omega_c=u/a$.\\
For a static QPC ($\omega=0$) the backscattering currents are ($\beta\omega_c\gg 1$)
\begin{eqnarray}\label{eq:Ist}
\langle I_p^{(\omega=0)}(V_c)\rangle&=&-\frac{4e^*\left |\tilde{\lambda}_p\right |^2}{\omega_c}\frac{e^*}{q_p} \sinh\left (\frac{\beta q_pV_c}{2}\right )\left (\frac{2\pi}{\beta\omega_c}\right )^{2\Delta_p-1}\nonumber\\
&\times&\mathcal{B}\left (\Delta_p-i\frac{\beta q_pV_c}{2\pi},\Delta_p+i\frac{\beta q_pV_c}{2\pi}\right ),
\end{eqnarray}
where $\mathcal{B}(x,y)$ is the Euler beta function and
\begin{eqnarray}\label{eq:scaling}
q_1&=&e^*, \ \ \ \ \ \ \Delta_1=\frac{\nu}{2}\left (K+\frac{1}{K}\right ),\nonumber\\
q_c&=&2e^*, \ \ \ \ \ \Delta_c=2\nu K,
\end{eqnarray}
are the backscattered charge and scaling dimensions for the processes.
Note that Eqs. \eqref{eq:Ist} and \eqref{eq:scaling} are valid both for quasiparticle tunneling in fractional TIs (with $\nu=1/(2n+1)<1$) and for electron tunneling in integer TIs (with $\nu=1$ and $e^*\equiv e$).
The smallest scaling dimension in Eq. \eqref{eq:scaling} identifies the dominant tunneling process. In particular one finds that single quasiparticle tunneling ($p=1$) dominates at weak interactions $K>K^*$, with $K^*=1/\sqrt{3}$. At stronger interactions $K<K^*$ the charge process ($p=c$) dominates.
In any case, for a static QPC, both processes always reduce the total current ($I_p^{(\omega=0)}<0$ for $V_c>0$), as can be argued \cite{ref1} from Eq. \eqref{eq:Ist}.\\
However, when an ac modulated ($\omega\neq 0$) QPC is considered, the backscattering contributions are no longer driven by the bias only, but the time dependent constriction comes into play.
This can be seen by writing Eqs. \eqref{eq:I1K} and \eqref{eq:IcK} as
\begin{equation}\label{eq:I-I}
\langle I_p^{(\omega)}(V_c)\rangle=\frac{1}{4}\sum_{\eta=\pm}\left\langle I_p^{(\omega=0)}\left (V_c+\eta\omega/q_p\right )\right\rangle,
\end{equation}
which shows a striking similarity with the photo-assisted current.
We find that the current is driven by two effective channels \cite{Makogon06} with voltages $V_c\pm \omega/q_p$, and the sign of $\langle I_p^{(\omega)}(V_c)\rangle$ now depends on how the two terms in Eq. \eqref{eq:I-I} combine.
Even if there is only one time dependent parameter (the tunneling amplitude), the backscattering current has been interpreted \cite{Feldman} as originated from a pumping mechanism, since the effect of gauging out the bias voltages is to give rise to the additional time dependent parameter necessary to generate a pumping current.
In this case, it is no longer obvious that the constriction leads to a decreasing of the current, but the time dependent QPC may be able to pump a current in the forward direction, leading to a global current enhancement \cite{Feldman, Makogon06, Makogon07, Agarwal07, Salvay09, Schmeltzer01}.
At low temperatures, one finds from Eqs. \eqref{eq:Ist} and \eqref{eq:I-I} the necessary condition for the process $p$ to enhance the current to be $\Delta_p<1/2$. Under this condition, the enhancement is expected for $V_c\apprle \omega/q_p$.
\begin{figure}[!ht]
\centering
\includegraphics[width=8cm,keepaspectratio]{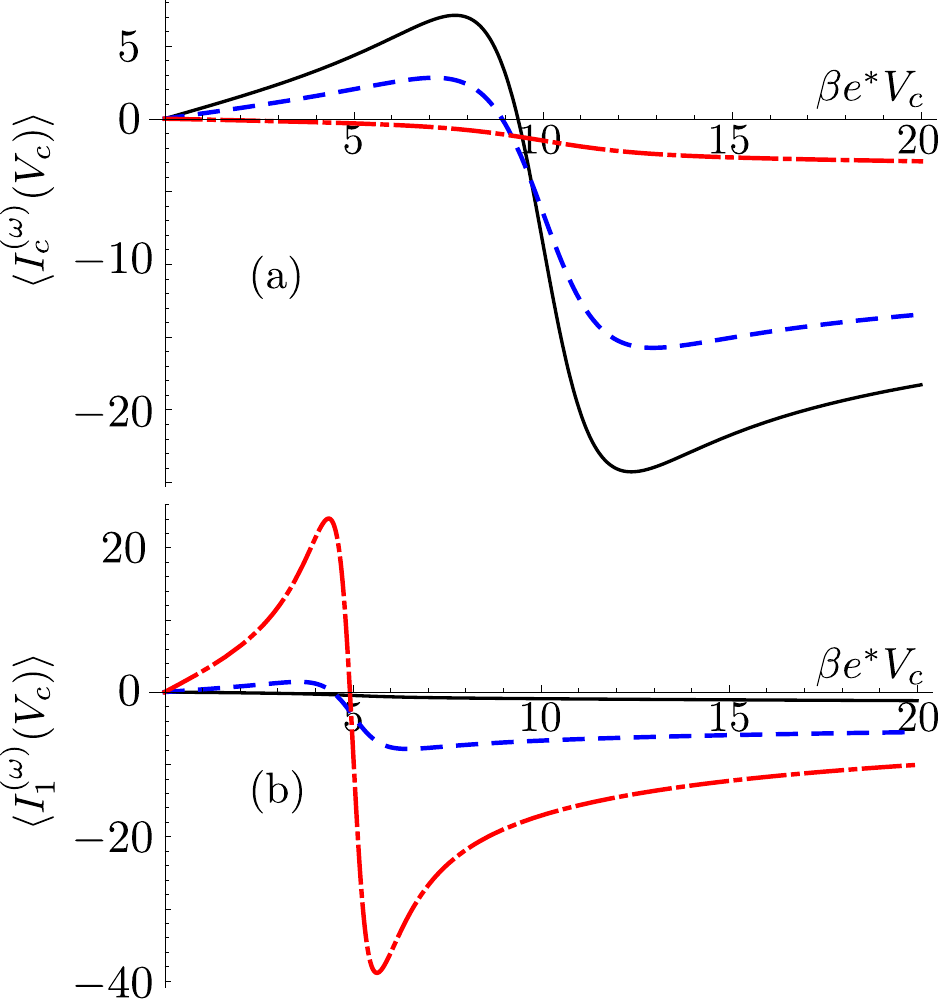}
\caption{(Color online) dc backscattering currents $\langle I^{(\omega)}_p(V_c)\rangle$ (units $e^*|\tilde{\lambda}_p|^2/\omega_c$) as a function of $\beta e^*V_c$ for a fractional TI with $\nu=1/3$, for $\beta\omega=10$, $\beta\omega_c=10^3$ and different interaction strength: $K=1$ (solid black), $K=1/\sqrt{3}\approx 0.577$ (dashed blue) and $K=0.3$ (dot-dashed red). (a) The single quasiparticle contribution enhances the total current at weak interactions for $V_c\apprle \omega/e^*$, while (b) the charge contribution enhances the total current at stronger interactions for $V_c\apprle \omega/(2e^*)$.}\label{fig4b}
\end{figure}
\\
Let us consider fractional TIs, and study whether the current is enhanced or decreased by the time dependent constriction.
Positive values of $\langle I_p^{(\omega)}(V_c)\rangle$ correspond to an enhancement of the current more than the universal value $I_c^{(0)}$ (remember we are focusing on $V_c>0$).
At weak interactions $K>K^*$ the dominant contribution is the single quasiparticle tunneling with scaling dimension $\Delta_1$ (see Eq. \eqref{eq:scaling}). In this regime $\Delta_1<1/2$ and the current is enhanced for $V_c\apprle\omega/e^*$ (Fig. \ref{fig4b}(a)).
The single quasiparticle process decreases the current for stronger interactions $K<\frac{1}{2\nu}\left (1-\sqrt{1-4\nu^2}\right )<K^*$ (red curve in Fig. \ref{fig4b}); however, in this regime tunneling is dominated by the charge process, which leads to enhancement already at moderate interactions $K<\frac{1}{4\nu}$ for $V_c\apprle\omega/(2e^*)$, as shown in Fig. \ref{fig4b}(b).
It is worth noticing that at the crossover $K\approx K^*$ both the single quasiparticle and charge processes contribute to enhance the current.
We thus expect a \emph{global enhancement} of the current in fractional TIs in the presence of a time dependent constriction, due either to single quasiparticle (for weak interaction) or to charge tunneling (for strong interaction).\\
Let us now compare this scenario with the one obtained for an integer TI, where only electrons can tunnel.
From Eq. \eqref{eq:scaling} we find that single electron tunneling, which dominates at weak interactions $K>K^*$, always decreases the current. At stronger interactions $K<K^*$ charge tunneling dominates, which however still leads to a decreasing as long as $K>1/4$.\\
We then conclude that, if interactions are not too strong $K>1/4$, the enhancement of the current in the presence of an ac modulated QPC is a peculiar feature of fractional TIs.\\
Other signatures can help to discriminate fractional TIs also for $K<1/4$.
Indeed, it is worth noticing that, as shown in Fig. \ref{fig4b}, a transition of $\langle I_p^{(\omega)}(V_c)\rangle$ from positive to negative values is expected around $V_c\approx\omega/q_p$, and thus depends on the backscattering process through the backscattered charge $q_p$. Since $V_c$ and $\omega$ can be externally controlled, one should be able to extract information about the backscattered charge, and thus on the fractional or integer nature of the system,  simply by varying the bias.
In particular, by observing whether the current displays a transition around $V_c\approx\omega/e^*$ or $V_c\approx\omega/(2e^*)$, one could also discriminate between single-particle and multi-particle tunneling.
In other words, the position of the transition in the backscattering current has the power to resolve the presence of different fractional excitations. In this perspective, this quantity has similarities with the photo-assisted differential conductance or finite frequency noise \cite{Safi10, Carrega12, Ferraro14}, since they are similarly affected by the scaling dimensions and by the Josephson resonances of the tunneling excitations.\\
Furthermore, the current enhancement represents a signature of quasiparticle (and \emph{not} electron) tunneling.
Indeed, evaluating the tunneling contribution of electrons (despite not the dominant process) would essentially result in replacing $e^*\to e$ and $\nu\to 1/\nu$ in Eqs. \eqref{eq:Ist} and \eqref{eq:scaling} and one can conclude that both single and double electron tunneling decrease the current for not too strong interactions (as long as $K>\nu/4$).\\
We have thus proposed a simple strategy to identify quasiparticle tunneling in fractional TIs.
This method could complement other peculiar responses, such as the quantized conductance, the stability to TR breaking perturbations or the power-law behavior of the backscattering current.
However we stress that, to identify quasiparticle tunneling, one does not need to study the details of the current (like for example for extracting the exponents of the power laws \cite{Beri12}), but the enhancement of the current in the presence of a time dependent QPC provides by itself a powerful signature of the presence of a fractional TI.
For strong interactions $K<1/4$, the current is enhanced both in fractional and in integer TIs; however, the transition from positive to negative values depends on the backscattered charge, allowing to discriminate between fractional and integer excitations. So we can conclude that the proposed protocol is stable against electron interactions.\\
Before concluding, we briefly comment on a different bias voltage configuration, that is the so called spin configuration, with $V_c=0$ and $V_s\neq 0$, corresponding to the presence of a bias voltage between the upper (at voltage $V_s/2$) and the lower edge (at voltage $-V_s/2$).
Here a net charge tunneling current flows across the constriction from one edge to the other, arising from a single quasiparticle tunneling contribution, given by Eq. \eqref{eq:I1K} with $V_c\to V_s$, and a spin tunneling contribution, given by Eq. \eqref{eq:IcK} with $V_c\to V_s$ and $K\to 1/K$.
The dominant contribution is the single quasiparticle one with scaling dimension $\Delta_1$.
In the presence of a static QPC the tunneling current always flows from positive to negative voltage, as expected.
However, when considering a time dependent QPC in fractional TIs the tunneling current may reverse its direction, flowing \emph{against} the bias.
This anomalous response, which is expected to occur as long as interactions are not too strong ($K>\frac{1}{2\nu}\left (1-\sqrt{1-4\nu^2}\right )$), represents another peculiar feature of fractional TIs with respect to their integer version, where it is not expected to occur at all.

\section{Conclusions}\label{conclusions}
We showed that an enhancement of the current is expected for a fractional topological insulator with a time dependent constriction, allowing for quasiparticle tunneling from one edge to the other.
This effect is not expected to occur neither in integer topological insulators (for not too strong interactions) nor in fractional topological insulators when electrons (and not quasiparticles) tunnel through the constriction, and thus represents an important tool to identify fractional quasiparticles.
This simple method allows to unambiguously identify fractional topological insulators without requiring the knowledge of the detailed form of the current (like for example for extracting the power law exponents).
Furthermore we showed that this effect is robust against electron interaction, and discussed how one can discriminate between single and multi particle processes.

\section*{Acknowledgments}

The supports of EU-FP7 via Grant No. ITN-2008-234970 NANOCTM and MIUR-FIRB2012 - Project HybridNanoDev via Grant No.RBFR1236VV are acknowledged.


\begin{thebibliography}{10}

\bibitem{vonKlitzing80} K. von Klitzing, G. Dorda, M. Pepper, Phys. Rev. Lett. \textbf{45}, 494 (1980).
\bibitem{Tsui82} D. C. Tsui, H. L. Stormer, and A. C. Gossard, Phys. Rev. Lett. \textbf{48}, 1559 (1982).
\bibitem{Laughlin83} R. B. Laughlin, Phys. Rev. Lett. \textbf{50}, 1395 (1983).
\bibitem{Haldane88} F. D. M. Haldane, Phys. Rev. Lett. \textbf{61}, 2015 (1988).
\bibitem{Kane05} C. L. Kane, and E. J. Mele, Phys. Rev. Lett. \textbf{95}, 226801 (2005).
\bibitem{Hasan10} M. Z. Hasan and C. L. Kane, Rev. Mod. Phys. \textbf{82}, 3045 (2010).
\bibitem{Qi11} X.-L. Qi and S.-C. Zhang, Rev. Mod. Phys. \textbf{83}, 1057 (2011).
\bibitem{BernevigNat06} B. A. Bernevig, T. L. Hughes, S.-C. Zhang, Science \textbf{314}, 1757 (2006).
\bibitem{Konig07} M. K\"{o}nig, S. Weidmann, C. Br\"{u}ne, A. Roth, H. Buhmann, L. W. Molenkampf, X.-L. Qi, and S.-C. Zhang, Science \textbf{318}, 766 (2007).
\bibitem{Roth09} A. Roth, C. Br\"{u}ne, H. Buhmann, L. W. Molenkamp, J. Maciejko, X.-L. Qi, and S.-C. Zhang, Science \textbf{325}, 294 (2009).
\bibitem{BernevigPRL06} C. Wu, B. A. Bernevig, and S.-C. Zhang, Phys. Rev. Lett. \textbf{96}, 106401 (2006).
\bibitem{Levin09} M. Levin and A. Stern, Phys. Rev. Lett. \textbf{103}, 196803 (2009).
\bibitem{Levin12} M. Levin and A. Stern, Phys. Rev. B \textbf{86}, 115131 (2012).
\bibitem{NeupertPRB11} T. Neupert, L. Santos, S. Ryu, C. Chamon, and C. Mudry,
Phys. Rev. B \textbf{84}, 165107 (2011).
\bibitem{NeupertPRL12} T. Neupert, L. Santos, S. Ryu, C. Chamon, and C. Mudry, Phys. Rev. Lett. \textbf{108}, 046806 (2012).
\bibitem{Li14} Wei Li, D. N. Sheng, C. S. Ting, and Yan Chen, arXiv:1407.1563.
\bibitem{Lindner12} N. H. Lindner, E. Berg, G. Refael, and A. Stern, Phys. Rev. X \textbf{2}, 041002 (2012).
\bibitem{Cheng12} M. Cheng, Phys. Rev. B \textbf{86}, 195126 (2012).
\bibitem{Burrello13} M. Burrello, B. van Heck, and E. Cobanera, Phys. Rev. B \textbf{87}, 195422 (2013).
\bibitem{Nikolic13} P. Nikolic, T. Duric, and Z. Tesanovic, Phys. Rev. Lett. \textbf{110}, 176804 (2013).
\bibitem{Beri12} B. Beri and N. R. Cooper, Phys. Rev. Lett. \textbf{108}, 206804 (2012).
\bibitem{Teo09} J. C. Y. Teo and C. L. Kane, Phys. Rev. B \textbf{79}, 235321 (2009).
\bibitem{Hou09} C.-Y. Hou, E.-A. Kim, and C. Chamon, Phys. Rev. Lett. \textbf{102}, 076602 (2009).
\bibitem{Dolcetto12} G. Dolcetto, S. Barbarino, D. Ferraro, N. Magnoli, and M. Sassetti, Phys. Rev. B \textbf{85}, 195138 (2012).
\bibitem{Schmidt11} T. L. Schmidt, Phys. Rev. Lett. \textbf{107}, 096602 (2011).
\bibitem{Liu11} C. -X. Liu, J. C. Budich, P. Recher, and B. Trauzettel, Phys. Rev. B \textbf{83}, 035407 (2011).
\bibitem{Strom09} A. Strom and H. Johannesson, Phys. Rev. Lett. \textbf{102}, 096806 (2009).
\bibitem{dePicciotto97} R. de Picciotto, M. Reznikov, M. Heiblum, V. Umansky, G.
Bunin, and D. Mahalu, Nature \textbf{389}, 162 (1997).
\bibitem{Saminadayar97} L. Saminadayar, D. C. Glattli, Y. Jin, and B. Etienne, Phys. Rev. Lett. \textbf{79}, 2526 (1997).
\bibitem{Chung03} Y. C. Chung, M. Heiblum, and V. Umansky, Phys. Rev. Lett.
\textbf{91}, 216804 (2003).
\bibitem{Dolev08} M. Dolev, M. Heiblum, V. Umansky, A. Stern, and D. Mahalu,
Nature \textbf{452}, 829 (2008).
\bibitem{Carrega11} M. Carrega, D. Ferraro, A. Braggio, N. Magnoli, and M. Sassetti, Phys. Rev. Lett. \textbf{107}, 146404 (2011).
\bibitem{Huang13} C.-W. Huang, S. T. Carr, D. Gutman, E. Shimshoni, and A. D. Mirlin, Phys. Rev. B \textbf{88}, 125134 (2013).
\bibitem{Giamarchi} T. Giamarchi, \textit{Quantum Physics in One Dimension} (Oxford University Press, Oxford, 2003).
\bibitem{ref2} Note that when talking about interactions we refer, in analogy with Ref. \onlinecite{Beri12}, to interactions on the edges given by Eqs. \eqref{eq:Hee_f} and \eqref{eq:Hee_b}, and not to the Coulomb interaction which gives rise to the fractional state as generated by the topological order in the bulk.
\bibitem{Feldman} D. E. Feldman and Y. Gefen, Phys. Rev. B \textbf{67}, 115337 (2003).
\bibitem{Souquet12} J.-R. Souquet and P. Simon, Phys. Rev. B \textbf{86}, 161410(R) (2012).
\bibitem{Lee12} Y.-W. Lee, Y.-L. Lee, and C.-H. Chung, Phys. Rev. B \textbf{86}, 235121 (2012).
\bibitem{Martin} T. Martin, \textit{in Nanophysics: Coherence and Transport, Les Houches Session LXXXI}, edited H. Bouchiat \textit{et al}. (Elsevier, Amsterdam,
2005).
\bibitem{ref1} Note that $\mathrm{sign}\{\langle I_p^{(\omega)}\rangle\}=-\mathrm{sign}\{V_c\}$, while $\mathrm{sign}\{I_c^{(0)}\}=\mathrm{sign}\{V_c\}$.
\bibitem{Makogon06} D. Makogon, V. Juricic, and C. M. Smith, Phys. Rev. B \textbf{74}, 165334 (2006).
\bibitem{Makogon07} D. Makogon, V. Juricic, and C. M. Smith, Phys. Rev. B \textbf{75}, 045345 (2007).
\bibitem{Agarwal07} A. Agarwal and D. Sen, Phys. Rev. \textbf{76}, 035308 (2007).
\bibitem{Salvay09} M. J. Salvay, Phys. Rev. B \textbf{79}, 235405 (2009).
\bibitem{Schmeltzer01} D. Schmeltzer, Phys. Rev. B, \textbf{63}, 125332 (2001).
\bibitem{Safi10} I. Safi and E. V. Sukhorukov, Europhys. Lett. \textbf{91}, 67008 (2010).
\bibitem{Carrega12} M. Carrega, D. Ferraro, A. Braggio, N. Magnoli, and M. Sassetti, New J. Phys. \textbf{14}, 023017 (2012).
\bibitem{Ferraro14} D. Ferraro, M. Carrega, A. Braggio, and M. Sassetti, New J. Phys. \textbf{16}, 043018 (2014).





\end{thebibliography}
\end{document}